\documentclass[runningheads]{llncs}
\usepackage[T1]{fontenc}
\usepackage{graphicx}
\usepackage{amsmath,amssymb,amsfonts}
\usepackage{algorithmic}
\usepackage{algorithm2e}
\usepackage{textcomp}
\usepackage{xcolor}
\usepackage{url}
\usepackage{hyperref}
\usepackage{booktabs}
\usepackage{caption}
\usepackage{subcaption}
\begin{document}
\title{Making Social Platforms Accessible: Emotion-Aware Speech Generation with Integrated Text Analysis}
\titlerunning{Emotion-Aware TTS}
%
\author{Suparna De\inst{1}\orcidID{0000-0001-7439-6077} \and
Ionut Bostan\inst{2}\orcidID{0009-0003-8656-0454} \and
Nishanth Sastry\inst{1}\orcidID{0000-0002-4053-0386}}
\authorrunning{S. De et al.}
%
\institute{School of Computer Science and Electronic Engineering, 
University of Surrey, Guildford UK \and
Nquiringminds, University Pkwy. Southampton, UK SO16 7PT \\
\email{\{s.de,n.sastry\}@surrey.ac.uk}\\ 
\email{ionut@nquiringminds.com}}
%
\maketitle              
\begin{abstract}
Recent studies have outlined the accessibility challenges faced by blind or visually impaired, and less-literate people, in interacting with social networks, in-spite of facilitating technologies such as monotone text-to-speech (TTS) screen readers and audio narration of visual elements such as emojis. Emotional speech generation traditionally relies on human input of the expected emotion together with the text to synthesise, with additional challenges around data simplification (causing information loss) and duration inaccuracy, leading to lack of expressive emotional rendering. In real-life communications, the duration of phonemes can vary since the same sentence might be spoken in a variety of ways depending on the speakers' emotional states or accents (referred to as the one-to-many problem of text to speech generation). As a result, an advanced voice synthesis system is required to account for this unpredictability. We propose an end-to-end context-aware Text-to-Speech (TTS) synthesis system that derives the conveyed emotion from text input and synthesises audio that focuses on emotions and speaker features for natural and expressive speech, integrating advanced natural language processing (NLP) and speech synthesis techniques for real-time applications. 

Our system also showcases competitive inference time performance when benchmarked against the state-of-the-art TTS models, making it suitable for real-time accessibility applications.

\end{abstract}
\section{Introduction}
Making social networks accessible to vulnerable populations such as the blind and visually impaired (BVIP), especially in low-income settings, or the less-literate, relies on innovations such as screen readers with text-to-speech (TTS)~\cite{podsiado16_interspeech}, 
and audio description (AD) of creative media and emojis in text \cite{Sridharan2024,Tigwell2020}. 

The quality of synthetic voices and AD narrative text is more important to this user group than any other, as they rely on TTS for most of their interactions with devices~\cite{podsiado16_interspeech}.
\par
There are efforts to make social networks more accessible for BVIP and less-literate populations, 
with interactive voice-based social media platforms  \cite{gramvaani2016,Vashistha2015}
TTS narrations for emojis~\cite{Tigwell2020}, and software to add emotional markup tags into children's stories to enable story-telling software for BVIP children~\cite{Nicholls2000}. Traditional TTS practices suffer from an unnatural, monotone and ``robotic" quality of speech, meaning that the experience of social networks is hampered for them, since it is the ``speaker's voice that transmits the meaning and the affective dimension''~\cite{Sridharan2024}. Thus, there is a need for emotion-aware TTS for achieving more accessible social networks.

Data-driven approaches for TTS, such as concatenative synthesis~\cite{bulut2002expressive}~\cite{bulyko2001joint}~\cite{hirai2004using} (stitching natural pre-recorded speech sounds to form new words and phrases~\cite{indumathi2012survey}), and unit selection synthesis~\cite{black1997automatically,conkie1999robust}(storing multiple instances of each unit with varying prosodies in an inventory), allow for natural-sounding speech, but are, however, dependent on access to large speech fragments corpora, generating only words or phrases already present in the vocabulary and lack the flexibility to adjust to different emotions expressed in real-time input text \cite{tan2021survey}. Additionally, social media text features abbreviations and linguistic slang, which make automated emotion-aware TTS for such text more challenging.
\par
The generation of natural-sounding speech is challenging as it necessitates accurately capturing the nuances of human speech, including pitch, tone, and rhythm. Recent advances such as Tacotron2 \cite{shen2018natural} and FastSpeech \cite{ren2019fastspeech}, have led to the development of speech synthesis systems with appreciable quality that comparably resemble natural speech \cite{fastSpeech2,Nithin2022}.
Existing approaches for emotional TTS such as one-hot representation of the emotion label \cite{lee2017} or global style tokens-based frameworks \cite{wang2018gst} require manual input of the appropriate emotion for text, limiting their scalability and generalisability to new scenarios. The lack of high-quality and emotion-aware audio data, e.g. being free from background noise, and alignment between emotional tones or styles with the textual context or intention, makes it difficult for machine learning (ML) algorithms to accurately capture the subtle nuances of emotions in order to generate emotion-aware audio. Recent advances in transformer-based models, such as Bidirectional encoder representations for transformers (BERT) \cite{devlin2018pretraining} enable combining a transformer model for emotion prediction with a speech synthesis model, such as that in \cite{mukherjee22_interspeech}, which combines a text classifier with a voice synthesis model. 

We extend this line of research to multi-speaker (both male and female) and multiple emotion scenarios by creating a system for generating natural-sounding emotional speech. Extending the number of emotions considered as well as the speaker features brings new challenges for a TTS model as ``emotion information is affected by various paralinguistic characteristics of speech such as pitch, tone and tempo" \cite{emoqtts}. We address these challenges by training two separate models: one for text analysis and one for voice synthesis. The emotion embeddings acquired during the text model training are employed during the audio model inference to generate emotional speech automatically. The transformer-based architecture for emotion prediction accurately captures the subtle nuances of various emotions, while the adaptations we propose to existing voice synthesis systems (FastSpeech2 \cite{fastSpeech2}) enable the system to work with multiple speakers' characteristics and multiple emotion voice data. This enables the generation of emotionally expressive speech congruent with the emotional content of the input text without manual intervention. 
\par

Our model introduces new capabilities that empower users to select both speaker's identity and the emotional tone, hence filling gaps in current TTS systems. For instance, the method presented in \cite{Zhang2023AudiobookSW} uses BERT embedding and text-predicted prosody control, however, it is still unable to grant users the control over the voice and emotion. Chandra et al. \cite{chandra2024exploringspeechstylespaces} introduce the language model-based approach without any labeled data, however users cannot manipulate the voice output characteristics directly. By bringing the option of speech and emotion to the user, we open the door to personalisation in digital assistive technologies like screen reader software to make social networks more accessible.

A focus of our investigation lies in handling the emotional aspect of speech. We introduce an innovative technique of using an Emotion Embedding Transformation, which helps our model understand and recreate subtle emotional nuances more accurately in synthesised speech.
Compared to the state-of-the-art, our main contributions lie in considering an increased variety of emotions and ability to discern emotional nuances for a wider range of speaker demographics, to capture a broader range of emotional experiences, enabling reproducing real-world speech patterns and local languages. Our model is trained on the GoEmotions~\cite{demszky2020goemotions} database of Reddit comments, thus containing text that is representative of the linguistic features of social media text. We also ensure a balanced representation of speaker samples, with male and female voice data incorporated to generate synthesised speech for a wide range of applications. 
We evaluate our proposed end-to-end system on objective measures of inference speed and Root Mean Square Error (RMSE). Our developed system achieves comparable inference time performance against state-of-the-art TTS models such as Natural Speech \cite{tan2022naturalspeech}, Glow-TTS \cite{glowtts}, Grad-TTS \cite{gradtts} and VITS \cite{vits}. 
Our work contributes to community efforts with the developed models made publicly available: \url{https://bit.ly/4fa81NH} and a proof-of-concept demo: \url{https://bit.ly/3YgIE6L}.

\section{Related Work}

\subsection{Background}
Depending on the model architecture, the TTS process involves a few main components: a text analysis frontend module, an acoustic module, a spectrum generator and an audio rendering module. The grapheme-to-phoneme (G2P) conversion transforms a series of characters into a sequence of phonetic symbols, transforming a character string, such as ``author," into its corresponding phonetic representation: [AO, TH, ER]. The text analysis module performs text normalisation by breaking down the input text into its parts, such as words, phrases, and punctuation. The acoustic module takes in the G2P conversion and, together with the spectrum generator, converts this information into a mel spectrogram, which is a visual representation of an audio signal's frequency content over time, with the frequencies converted to the mel scale. The mel scale more accurately reflects the human perception of pitch differences in audio signals, especially at lower frequencies. Finally, the audio rendering module, the vocoder, converts the mel spectrogram into speech.

\subsection{Emotional speech synthesis}
Traditional concatenative emotion-aware speech synthesisers \cite{bulut2002expressive,bulyko2001joint,hirai2004using} work by concatenating speech fragments stored in databases, based on the input text. The limitation of these systems is that they are only able to generate words or phrases already present in the vocabulary, whereas our proposed approach generalises well on unseen text, meaning that it is more flexible and not bound by the data it has been trained on.
\par
Deep learning approaches such as WaveNet~\cite{oord2016wavenet} employ a generative model composed of a stack of convolutional layers without pooling, maintaining the same time dimensionality between input and output. It operates directly on raw audio waveforms, utilising softmax distributions to model the conditional distributions over individual audio samples.
Tacotron~\cite{wang2017tacotron} uses an attention-based seq2seq model, with the attention-based decoder enabling content-based alignment, while the encoder is built to extract robust sequential text representations. Its successor, Tacotron 2~\cite{shen2018natural}, consists of a recurrent seq2seq feature prediction network with attention and a modified WaveNet vocoder, resulting in higher-quality audio generation and a more accurate text-to-speech system.
The non-autoregressive TTS model FastSpeech~\cite{ren2019fastspeech}, has been identified as one of the most successful models due to its use of knowledge distillation, and introducing duration information, to expand the text sequence to match the length of the mel-spectrogram sequence. However, it suffers from a complicated training process and information loss in target mel-spectrograms compared to the ground-truth ones. 
To address these challenges and better handle the one-to-many mapping problem of one text input mapped to multiple speech outputs depending on the emotion context, FastSpeech2~\cite{fastSpeech2} (FS2) was introduced in 2020. Since then, numerous research efforts~\cite{chien2021investigating,emovie,mukherjee22_interspeech} have utilised its model, highlighting the growing importance of emotional TTS.
\par

Recent research includes EMOVIE~\cite{emovie}, which introduces a new dataset of Mandarin Chinese emotional speech and a simple emotional TTS model based on FS2 to control the emotion of generated speech, with an emotion predictor (Bi-LSTM classifier model) to encode the emotion label to a latent emotion embedding, and an emotion controller.
In comparison, we employ a transformer-based emotion classification model, which is better able to capture the context in the input text. A slightly different approach, using a limited amount of speech data \cite{zhou2021}, consists of a seq2seq emotional voice conversion (EVC) framework that generates the emotion embeddings from a set of reference utterances to generate the converted acoustic features. The proposed model is trained to perform both EVC and emotional TTS; however, EVC is the primary focus. Addressing the monotonous expression of synthesised speech is the focus of \cite{emoqtts}, which considers fine-grained emotion intensity at the phoneme-level by distance-based intensity quantization, by considering how far the vector is from the centroid of the neutral emotion. Such an approach is dependent upon the audio training dataset having a substantial word frequency spectrum with a variety of phonemes, including for words which have difficult pronunciations. Moreover, this approach also removes the speaker information from the emotion embedding.
Another method \cite{mukherjee22_interspeech} based on the FS2 neural TTS architecture, utilises a BERT-based text modelling approach to develop a text-aware emotional TTS system with BERT embeddings to capture the fluctuations in emotional expression within the text directly. Different from existing research on emotion detection using a two-stage training system \cite{chandra2024exploringspeechstylespaces}, our process employs a faster approach with RoBERTa for emotion detection, with the non-verbal emotional input enabling additional quality improvements. The fact that our model can be run in real-time inference allows its integration in assistive technologies such as screen readers and AD creative media elements, enabling its application in accesible voice-based social networks.
Additionally, there is currently no known public codebase for showcasing an English emotional TTS pipeline, a challenge that this paper also aims to address.

\section{Datasets}
\subsection{EmoV DB Dataset}
We use the EmoV-DB \cite{adigwe2018emotional} database to build models to produce and control emotional speech. The data was collected from English and French speakers (only English recordings were used in our work) in anechoic chambers, with the recordings segmented manually, to ensure high-quality recordings and sufficient diversity. 
The audio was recorded by four native English speakers, including two females and two males. The spoken sentences were sourced from the CMU-arctic \cite{Kominek2004TheCA} database, and the speakers spoke the utterances in four different emotions: amusement, anger, sleepiness, and disgust, and a neutral state. 
The data was initially recorded at a sample rate of 44.1kHz but was subsequently downsampled to 16kHz and saved in the 16-bit PCM WAV format.

\subsection{GoEmotions Dataset}
The GoEmotions database was selected for emotion classification from text. This database is one of the largest human-annotated datasets, containing 58,000 Reddit comments labelled for 27 emotions categories or `neutral'.
To use the emotion embeddings from the text classifier as input to the TTS model's encoder during inference, the sentiment embeddings in the classification model must match the ones in the voice synthesis model. Therefore, only four sentiments plus neutral were selected from the dataset: ``amusement", ``anger", ``disgust", ``neutral", and ``sleepy".
As a result, the raw version of the GoEmotions dataset, made up of over 210,000 sentences, was used instead. However, the resulting neutral sentiment had over 50,000 sentences, requiring the removal of 60\% of the neutral comments to balance the data. 

\subsection{Audio pre-processing}
In order to generate natural-sounding speech, the audio and text data must be synchronised. To achieve this, an external speech processing tool, the Montreal Forced Aligner (MFA)~\cite{mcauliffe17_interspeech}, is used, which uses text transcription and audio files to estimate the alignment between text and sound. A TextGrid file is produced as an output of this process which consists of words and phonemes. Each word is then aligned to its corresponding audio segment. It also provides information about the timing and duration of each phoneme in the audio file.
As an illustration, consider the word ``author". Based on the EmoV DB dataset, the word can be broken down into the following phonemes: A01, TH, ER0. The TextGrid to IntervalTier conversion represents the phonemes by intervals with xmin and xmax values, spanning from 0 to 1.06 seconds, where 1.06 seconds is a person's total time to speak all the phonemes. For instance, the intervals assigned to the phonemes: A01, TH, ER0 are A01: $xmin$: 0, $xmax$: 0.66, TH: $xmin$: 0.66, $xmax$: 0.94, and ER0: $xmin$: 0.94, $xmax$: 1.06.

\section{Emotion-Aware TTS Pipeline}
\begin{figure}[ht]
  \centering
  \includegraphics[width=0.5\linewidth]{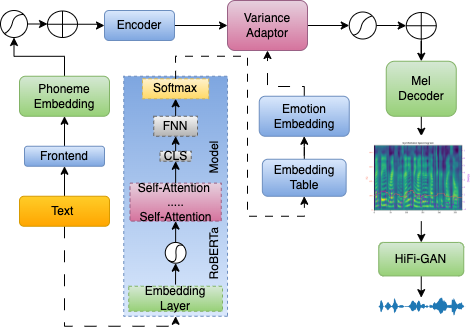}
  \caption{Emotion-Aware TTS Pipeline
  \label{fig:pipeline}}
\end{figure}

Our end-to-end pipeline, shown in Figure~\ref{fig:pipeline}, is based on FastSpeech2, a non-auto regressive TTS model, and has a number of components: a frontend that normalises the text and also includes a grapheme-to-phoneme model, the acoustic model, consisting of the encoder, variance adaptor and mel decoder. We employ HiFi-GAN~\cite{kong2020hifigan}, a Generative Adversarial Network (GAN) for efficient and high fidelity speech synthesis, to generate the waveforms.
The acoustic model includes speaker and emotion embeddings to handle multi-speaker and multi-emotion scenarios. The speaker embeddings are added to the output of the encoder, allowing the model to learn speaker-specific information. Similarly, the emotion embeddings are processed through an emotion linear layer and added to the encoder output, allowing the model to capture emotion-related information from the input text. To provide additional context for generating emotion-aware speech, emotion embeddings are incorporated into the model during training. This is achieved by adding a lookup table that stores the emotion-embeddings of a fixed dictionary and size.

\subsection{RoBERTa Emotion Classification Model}
Our approach utilises the RoBERTa~\cite{liu2019roberta} model to generate the representation of the input sentences, which is used for emotion classification. We chose RoBERTa for our model as it displayed better accuracy, and more stable and consistent F1-score and loss curves for a range of learning rates and batch sizes, compared to the BERT model. A smoother F1 score and loss curve indicate a more stable learning process, suggesting that the model is less likely to overfit or underfit the training data.
Our dataset is pre-processed using a tokenisation function, which formats the input sentences into a uniform format. Padding and truncation ensure a consistent sequence length of 128 tokens. We then feed this pre-processed data into RoBERTa, modified explicitly for sequence classification tasks. The model is designed with five output labels, each corresponding to our chosen emotion categories: ``amusement", ``anger", ``disgust", ``neutral", and ``sleepy". In this model, the final hidden states of a special token, known as [CLS], represent the entire text input. This representation, known as RoBERTa embedding, encodes key information for emotion classification. To prevent overfitting, the model's architecture comprises attention mechanisms and dropout layers. It also features layer normalisation and dense layers, which aid in capturing complex patterns within the dataset. Finally, the classification head of the RoBERTa model utilises a softmax function to generate probabilities for each emotion category that helps translate the high-dimensional learned features into final predictions, providing a distribution over our chosen emotion categories.

\subsection{Neural Emotion-aware TTS}
Pre-processing steps for TTS systems include text normalisation and grapheme-to-phoneme conversion.
For this, we standardise the input text in a form that the TTS model can effectively understand by using various methods such as ASCII transliteration, lowercase conversion, abbreviation expansion, number expansion, and whitespace collapse. Abbreviations such as `Dr.' are expanded to `doctor' and `St.' to `saint', using a predefined list of standard abbreviations and their corresponding expansions. Numbers are processed to replace numerals with their written form, e.g. `2' becomes `two'. Once properly normalised, the text is converted from graphemes (written characters) to phonemes (sound units) (G2P), by using an off-the-shelf module for G2P conversion of English text.
This transformation is necessary for accurate speech synthesis since the pronunciation of a word does not always directly align with its spelling. To prepare the audio data, the Montreal Forced Aligner (MFA) tool was used to generate the required TextGrid files containing essential phonemes and durations necessary for predicting the voice's duration, pitch and energy.

\subsubsection{Phoneme Embedding}
At this stage, information from the audio file name, TextGrid files, and .lab files is concatenated together and saved to a text file and separated by a pipe, as follows:
``neutral\_281-308\_0287|bea|\{K IY1 P AH0 N AY1 AA1 N HH IH1 M\}|Keep an eye on him.|neutral". The different field types are as follows:
\begin{itemize}
    \item File ID: \textbf{`neutral\_281-308\_0287'}
    \item Speaker ID: \textbf{`bea'}
    \item Phoneme sequence: \textbf{`\{K IY1 P AH0 N AY1 AA1 N HH IH1 M\}'}
    \item Text: \textbf{`Keep an eye on him.'}
    \item Emotion: \textbf{`neutral'}
\end{itemize}
This data is fed into the model:the first element, ``neutral\_281-308\_0287" serves as an identifier, while the name element, $bea$, is utilised for speaker embeddings. The phoneme sequence, {K IY1 P AH0 N AY1 AA1 N HH IH1 M}, is processed by the encoder and then passed to the variance adaptor to predict duration, pitch, and energy. Lastly, the emotion information is employed as input to the emotion embedding table, enhancing the model's ability to capture emotional nuances.

\subsubsection{Encoder}
The Encoder is made up of a combination of transformer and convolution components. It is primarily based on transformer design but also uses Fast Fourier Transform (FFT) blocks, which include a multi-head self-attention mechanism followed by a position-wise feed-forward network with gated convolution. The phoneme sequence is first tokenised and numerically represented. This sequence is then fed into the Encoder, which generates the hidden sequence.
To process the emotion embeddings before integrating them into the model, a sequential linear layer followed by a ReLU activation function is applied. This change allows the model to create a speech that better fits the context and conveys emotions effectively. The processed emotion embeddings are incorporated into the Encoder’s output, which is then passed through the variance adaptor.
For an input sequence of phonemes, $P = \{p_1, p_2, \dots, p_n\}$, where $n$ is the number of phonemes, we map each phoneme to its corresponding embedding using an embedding matrix $E \in \mathbb{R}^{(d \times V)}$, where $d$ is the dimension of the embedding space and $V$ is the vocabulary size (number of unique phonemes). The resulting phoneme embedding sequence is $X = \{x_1, x_2, \dots, x_n\}$, with $x_i = E \cdot p_i$.
\par
In this example, the input sequence of phonemes $P$ is represented as a sequence of 3-dimensional embeddings $X$, which can be further processed by the model.
The encoder processes the phoneme embedding sequence $X$ using a stack of self-attention layers and 1D-convolution layers. The output of the encoder, which is the phoneme hidden sequence $H$, can be represented as:
\vspace{-0.2cm}
\begin{equation}
    H = \text{Encoder}(X)
\end{equation}

The variance adaptor adds duration, pitch, and energy information to the phoneme hidden sequence $H$. We denote the log duration predictions as $D = \{d_1, d_2, \dots, d_n\}$, the pitch predictions as $P = \{p_1, p_2, \dots, p_n\}$, and the energy predictions as $E = \{e_1, e_2, \dots, e_n\}$.
The variance adaptor then expands the phoneme hidden sequence based on the predicted durations and adds the pitch and energy embeddings to create an adapted hidden sequence $H'$:
\vspace{-0.2cm}
\begin{equation}
    H' = \text{Variance\_Adaptor}(H, D, P, E)
\end{equation}
\noindent These predictions are obtained from their respective predictors:
\begin{align}
    D &= \text{Duration\_Predictor}(H) \\
    P &= \text{Pitch\_Predictor}(H) \\
    E &= \text{Energy\_Predictor}(H)
\end{align}

\subsubsection{Mel-spectrogram Decoder}
\begin{figure}
    \centering
    \includegraphics[width=\textwidth]{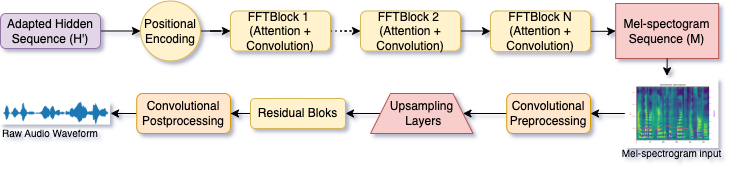}%
    \caption{Mel-spectrogram Sequence to Audio \label{fig:hifi}}%
\end{figure}

The mel-spectrogram decoder takes the adapted hidden sequence $H'$ and converts it into a mel-spectrogram sequence $M = \{m_1, m_2, \dots, m_t\}$, where $t$ is the number of time frames in the mel-spectrogram. The decoder uses a stack of self-attention layers and 1D-convolution layers (top half of Figure~\ref{fig:hifi}):
\vspace{-0.2cm}
\begin{equation}
    M = \text{MelSpectrogram\_Decoder}(H')
\end{equation}

During inference, HiFi-GAN\cite{kong2020hifigan} is used to convert mel-spectrograms into raw audio waveforms. The input mel-spectrogram is passed through a 1D convolutional layer for feature extraction and to adjust the number of channels. The features are then upsampled by a specific factor, gradually increasing the size of the feature map until the target audio length is reached. The upsampled feature map is then passed through a series of residual blocks. Each residual block contains two convolutional layers with different kernel sizes and dilations. The processed feature map goes through another 1D convolutional layer to generate the final raw audio waveform, which is then saved into an audio file (Figure~\ref{fig:hifi} bottom).
\subsubsection{Emotion Embedding}
During training, instead of using the predicted label of the text emotion classification model, the audio emotion label is captured from the audio file's name and is used as a conditional input to the emotional embedding table.
During training:
\begin{itemize}
    \item A \texttt{emotion} variable is crated as a NumPy array with a single element, \texttt{args.emotion\_id}, with a possible value from 0 to 4.
    \item If the \texttt{multi\_emotion} flag in the model configuration is set to True, the code reads the \texttt{emotions.json} file, which contains the mapping of emotions names to their corresponding IDs: \texttt{\{"amused": 0, "anger": 1, "disgust": 2, "neutral": 3, "sleepiness": 4\}} and calculates the number of unique emotions (n\_emotion) present in the dataset.
    \item An embedding layer \texttt{self.emotion\_emb} is created with size $=$ number of unique emotions \texttt{n\_emotion} and dimension $=$ encoder hidden size. This layer converts the emotions IDs into continuous vectors, for the model to process.
    \item The \texttt{self.emotion\_linear} layer consists of a linear transformation, fully connected layer, followed by ReLU activation function. This layer maps the emotion embedding vector to a new representation with the same size and dimension as the encoder hidden size \texttt{model\_config
    ["transformer"]
    ["encoder\_hidden"]}. 
\end{itemize}
During inference, the previously learned embedding table is used to map the predicted text emotion to a fixed size emotion embedding generated by the pre-trained RoBERTa-base model.

\subsubsection{Details of Training Parameters}

The transformer model configuration includes an encoder and a decoder, each with specified layers and hidden units. The encoder comprises 4 layers, each with 2 heads and 256 hidden units. The decoder configuration includes 6 layers, each with 2 heads and 256 hidden units. The model is designed to handle multi-emotion inputs with a multi-speaker setting. The variance predictor model uses filters of size 256 with a kernel size of 3, and a dropout rate of 0.5 for regularization.
The HiFi-GAN vocoder with a ``universal" speaker setting is employed. The optimizer is configured with a batch size of 16, and parameters for the Adam optimizer include beta-s set to 0.9 and 0.98, an epsilon of 1e-9, and a weight decay of 0.0. Gradient clipping is set at 1.0 with an accumulation step of 1.
For learning rate scheduling, a warm-up step of 4000 is used, and the annealing steps are set at 300K, 400K, and 500K with an anneal rate of 0.3.

\subsubsection{Model Optimisation}
The FastSpeech2 loss function is used to measure the differences between the model’s predicted output and actual target values (ground truth). Adam optimiser is used to adjust the parameters and minimise the loss during training. The loss function consists of multiple loss components: Mel-spectrogram loss, Postnet Mel-spectrogram loss, pitch loss, energy loss and durration loss. The Total loss is the sum of these individual losses. The mel-spectrogram and post mel-spectrogram losses utilise L1Loss to compute the Mean Absolute Error (MAE) between predicted and ground truth values:
\vspace{-0.2cm}
\begin{equation}
\text{L1Loss} = \frac{1}{N} \sum_{i=1}^{N} |y_{\text{pred}_i} - y_{\text{true}_i}|
\end{equation}

Pitch, energy and duration losses are computed with the help of Mean Square Error loss (MSE) which calculates the error between the predicted and ground truth values:
\vspace{-0.2cm}
\begin{equation}
\text{MSELoss} = \frac{1}{N} \sum_{i=1}^{N} (y_{\text{pred}_i} - y_{\text{true}_i})^2
\end{equation}

The total loss is a sum of the individual loss components:
\vspace{-0.2cm}
\begin{align}
    L_{total} &= L_{mel} + L_{postnet\_mel} + L_{pitch} + L_{energy} + L_{duration}
\end{align}

\subsection{TTS Pipeline Deployment}
We have deployed the developed speech synthesis model on the HuggingFace platform to allow users to generate synthetic speech samples through an interactive web interface, enabling users to generate emotion-aware speech samples by specifying various parameters, including input text, speaker ID, embedding type, and desired emotion, for the model to generate a corresponding audio file. Users can also interact with the model via an API through HTTP requests. 
It is important to note that the inference time may be slower than the results presented in this paper as the model operates on a CPU compared to the GPU used for testing.

\section{Evaluation}
The evaluation of the TTS system focuses on inference speed and Root Mean Square Error (RMSE). The inference speed is calculated for each speaker and compared. Also, the time required to generate one second of synthesised speech is benchmarked against other TTS models that utilise a similar HiFi-GAN vocoder. Since the EmoV DB dataset lacks ground truth mel spectrograms, it is not feasible to use Mel Cepstral Distortion (MCD) to measure the differences between the ground truth and generated spectrograms. Instead, RMSE compares the actual wave files against the generated ones. Lower error values indicate higher quality in the generated samples.
\subsection{Inference Speed Performance of Synthesised Speech}
Table~\ref{tab:full_time} shows the time taken to synthesize one text sample with manually added emotion embeddings compared to embeddings generated by the text classifier, using different speakers and emotions. Additionally, the table displays the number of words in each text sequence.
Considering the additional text processing, the results indicate that there is no drawback in using the generated embeddings during speech synthesis. In fact, a slight decrease in inference time can be observed in some instances.
\vspace{-0.2cm}
\begin{table}[ht]
\centering
\begin{tabular}{@{}lllllllr@{}}
\toprule
Method  & Speaker & Gender & Emotion & Sample ID & Time (s) & Words\\ 
\midrule
RoBERTa & Bea & female & amused & 0173 & 0.52 & 7\\
Manual & Bea & female & amused & 0173 & 0.55 & 7\\ 
\bottomrule
\end{tabular}%
\caption{Inference time comparison between manually selected emotions and generated ones. GPU Tesla T4 16GB RAM \label{tab:full_time}}
\end{table}

Table~\ref{tab:full_time_all} presents the time required to generate sound waves for two sentences with different emotions and speakers. Although speaker number three, Josh, lacks ``anger" and ``disgust" emotion samples in the dataset, the model could represent and synthesise these emotions for the speaker. However, there are no ground truth samples available for comparison results. Therefore only the sentiments ``amused" and ``neutral" are chosen for this test as these are the only ones present for all the speakers.
\vspace{-0.2cm}
\begin{table}[ht]
\centering
\begin{tabular}{@{}llllllr@{}}
\toprule
Speaker & Gender & Emotion & Sample ID & Time (s) & Words\\ \midrule
Bea & female & amused & 0173 & 0.55 & 7 \\
Bea & female & neutral & 0173 & 0.54 & 7\\ 
Jenie & female & amused & 0140 & 0.56 & 8\\
Jenie & female & neutral & 0173 & 0.73 & 7\\ 
Josh & male & amused & 0173 & 0.48 & 7\\
Josh & male & neutral & 0173 & 0.46 & 7\\ 
Sam & male & amused & 0173 & 0.58 & 7\\
Sam & male & neutral & 0173 & 0.47 & 7\\ 
\bottomrule
\end{tabular}
\caption{Inference time comparison for all speakers. GPU Tesla T4 16GB RAM \label{tab:full_time_all}}
\end{table}
\par
The results in Table~\ref{tab:full_time_all} show that the inference time is relatively consistent across different speakers and their emotional states. For example, the female speaker Bea has only minor differences in inference times when expressing ``amusement" or a ``neutral" emotion. This pattern is also observed for the other speakers, including Jenie, Josh, and Sam, whose inference times vary slightly across different emotional states. These results indicate that the TTS system is stable in generating speech for various speakers and emotions without significant differences in the synthesis of speech with different emotions.
Table~\ref{tab:seconds_time} compares the inference speeds of the Emotion-Aware TTS with those of previous TTS systems, as presented in  NaturalSpeech~\cite{tan2022naturalspeech}, as well as considering NaturalSpeech as an additional baseline. These results were measured using an NVIDIA V100 GPU. For a fairer comparison, our system was tested on an NVIDIA A100 GPU, since we did not have access to a V100 GPU. It is also important to note that the presented systems lack the capability to select different speakers or emotions, which adds to the model complexity (and hence, inference time) in our case.
\vspace{-0.2cm}
\begin{table}[htbp]
\centering
\begin{tabular}{@{}llllr@{}}
\toprule
System  & Vocoder & RTF & GPU\\ \midrule
NaturalSpeech & HiFi-GAN & 0.013 & V100 \\ 
Glow-TTS & HiFi-GAN & 0.021 & V100\\
Grad-TTS(100) & HiFi-GAN & 4.120 & V100\\ 
VITS & HiFi-GAN & 0.014 & V100\\
\midrule
\textbf{Emotional TTS} & HiFi-GAN & 0.080 & A100\\
\bottomrule
\end{tabular}
\caption{Inference speed comparison of RTF (real-time factor, time to synthesize 1 second waveform) for the neutral sentiment, compared to NaturalSpeech\label{tab:seconds_time} \protect\cite{tan2022naturalspeech}}
\end{table}
Considering the complexity of the Emotion-Aware TTS model, which includes a text classification model during inference, the 0.080-second inference time for each second of a synthesised waveform is quite reasonable. It is not far from the other systems presented, which directly generate the waveform from text, suggesting that it could be potentially utilised in real-time systems. 
\subsection{RMSE Analysis of Synthesised Speech}

Table~\ref{tab:rmse} compares the RMSE of synthesised and natural speech waveforms. 
\begin{table}
\centering
\begin{tabular}{@{}llllllr@{}}
\toprule
Speaker & Gender & Emotion & Sample & RMSE\\ \midrule
Bea & female & amused & 0173 & 0.1803\\
Bea & female & neutral & 0173 & 0.1683\\ 
Jenie & female & amused & 0173 & 0.0724\\
Jenie & female & neutral & 0140 & 0.1662\\
Josh & male & amused & 0173 & 0.1576\\
Josh & male & neutral & 0173 & 0.1769\\
Sam & male & amused & 0173 & 0.1290\\
Sam & male & neutral & 0173 & 0.1694\\
\bottomrule
\end{tabular}
\caption{Synthesised and natural speech waveforms comparison: RMSE\label{tab:rmse}}
\end{table}

\begin{figure}
     \centering
     \begin{subfigure}[b]{0.45\textwidth}
         \centering
         \includegraphics[width=\textwidth]{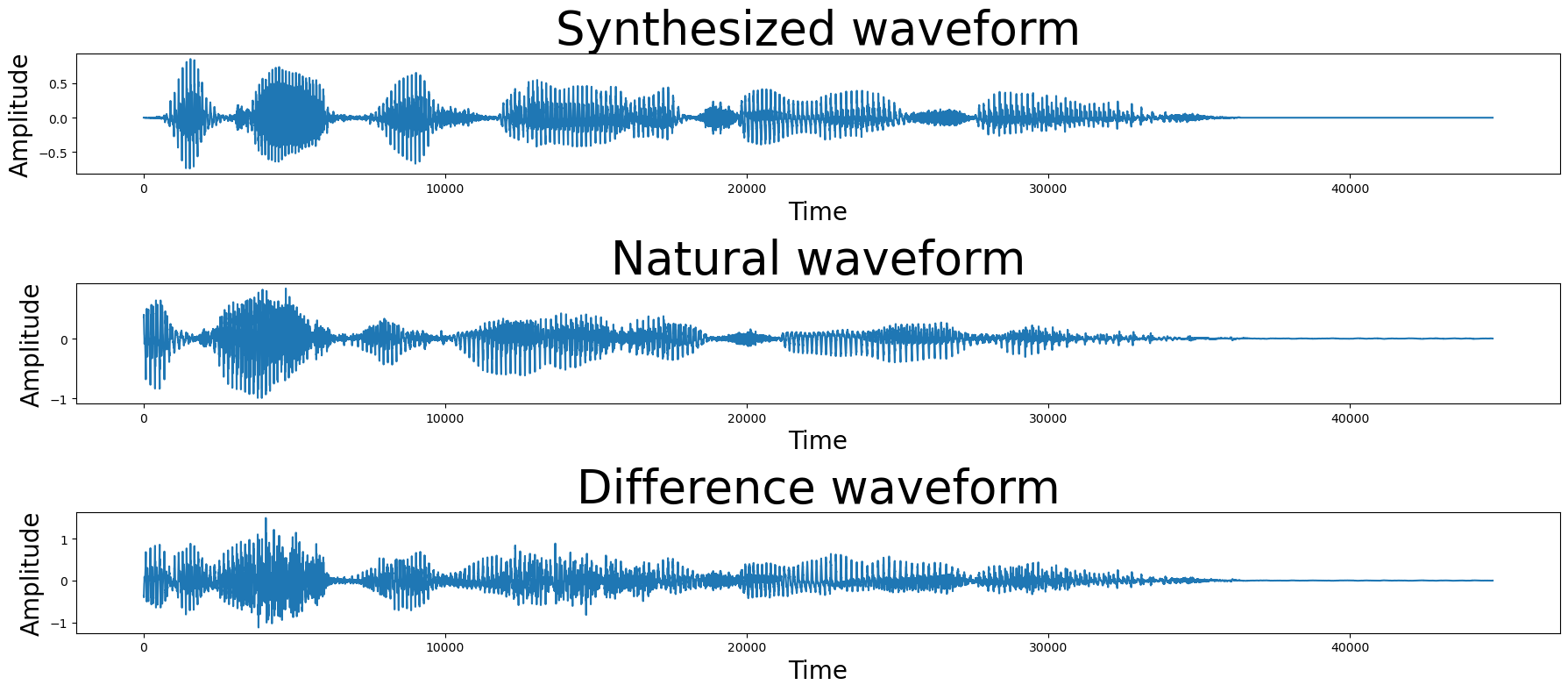}
         \caption{Josh 0173 `neutral'}
         \label{fig:gen1}
     \end{subfigure}
     \hfill
     \begin{subfigure}[b]{0.45\textwidth}
         \centering
         \includegraphics[width=\textwidth]{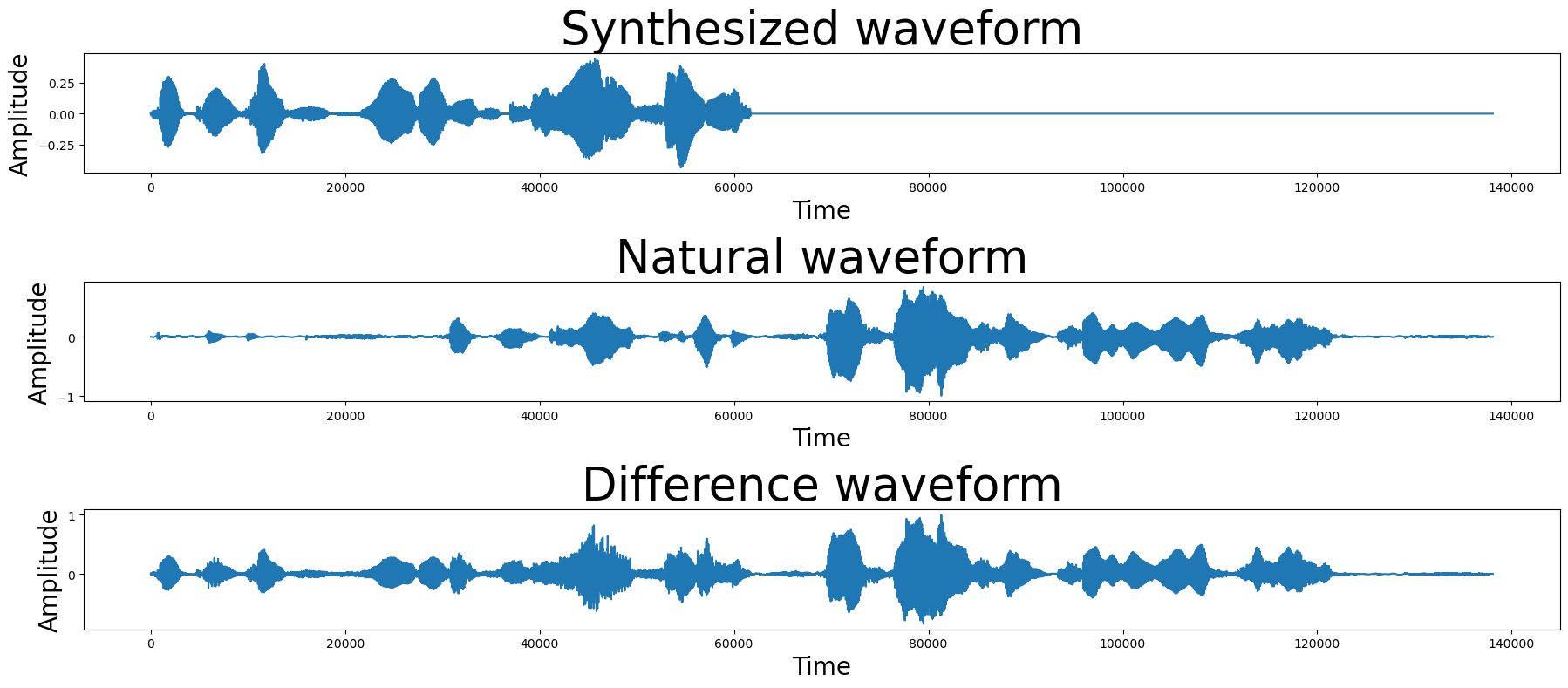}
         \caption{Sam 0173 `amused'}
         \label{fig:gen2}
     \end{subfigure}
     \hfill
        \caption{RMSE}
        \label{fig:2 graphs}
\end{figure}

The RMSE values presented in Table~\ref{tab:rmse} fall within an acceptable range, considering the dataset and the quality of the recordings. This error is better visualized in Figures~\ref{fig:gen1} and~\ref{fig:gen2}. To generate these images, some samples had to be padded due to differences in wavelengths; seen as a continuous line at the end of the synthesised or natural waveforms. 
All samples generated for these findings were produced with a 22,050 Hz sampling rate, the recommended sampling rate for TTS, as higher sampling rates make the data too voluminous for a neural network.

Figures~\ref{fig:gen1} and~\ref{fig:gen2} demonstrate that the model synthesizes the ``neutral" sentiment more accurately than the ``amused" sentiment, which is valid for all speakers. Further analysis of the RMSE images for different emotions and speakers (Figure~\ref{fig:gen2}) shows continuous lines in the natural waveform sample that indicate a pause at the beginning and also in-between words. After listening to the original audio file, it was confirmed that pauses were present at the start of the sample as well as between words. Further examination of other sentiment audio instances, such as ``sleepy" and ``anger," revealed pauses between words and other expressions not present in the text transcripts. For example, ``ha ha ha" was observed in the ``amused" samples, while ``yawning" was noted in the ``sleepy" examples. This is an important finding that could affect the model's overall performance.

\section{Conclusion}
In this paper, we introduce Emotion-Aware TTS, an approach to synthesise good-quality neutral and emotion-aware audio waveforms. The developed system demonstrates the potential for automatically generating emotion-aware speech audio by leveraging a text classifier, thus eliminating the need for manual intervention. By working seamlessly with the RoBERTa text classifier, the model can generalise effectively and synthesise audio waveforms from unseen text samples. Furthermore, the model showcases the capacity for selecting different speakers and emotions, which adds to its versatility, while still keeping competitive inference times, showing its potential for integration in assistive technologies for achieving accessible social platforms. 
Future directions of work include improving audio data quality and range, refining the accuracy of text classifiers, optimising the speech-generating methodology, and expanding the system's ability to express a wider range of emotions.

%
%
%
\bibliographystyle{splncs04}

\bibliography{emotiveTTS}

\end{document}